\documentclass[twoside,fleqn]{article}
\usepackage{espcrc2}
\usepackage{psfig}
\newcommand{\gev}{{\rm GeV}}
\newcommand{\bea}{\begin{eqnarray}}
\newcommand{\eea}{\end{eqnarray}}
\title{Heavy to light vector meson semileptonic decays 
\thanks{presented by Federico Mescia at ``Lattice 2002", Boston.}
\thanks{This work is supported in part by the European Network 
``Hadron Phenomenology and Lattice QCD", HPRN-CT-2000-00145.}}
\author{ {\tt SPQcdR} Collaboration\\ 
        A.~Abada\address[FR]{Universit\'e de Paris Sud, L.P.T.~(B\^at.~210),
	91405 Orsay-Cedex, France },
        D.~Becirevic\address[IT]{ Dip. di Fisica, Universit\`a ``La Sapienza", Piazzale Aldo Moro, I-00185 Rome, Italy},
        Ph.~Boucaud\addressmark[FR],
	J.M.~Flynn\address[UK]{Department of Physics and Astronomy,
        University of Southampton, Southampton SO17 1BJ,UK},
	J.P.~Leroy\addressmark[FR],
        V.~Lubicz\address[IT3]{Dip.di Fisica, Univ.di Roma Tre \& INFN-Sez.di         Roma III, Via
	della V. Navale 84, I-00146 Rome, Italy},
	F.~Mescia\addressmark[UK]}
\begin{document}

\begin{abstract}
New (preliminary) results for the form factors relevant for the semileptonic 
decays of heavy pseudoscalar to a light vector meson are presented. In 
particular, we discuss the form factors for $D\to K^\ast$ and 
$B\to \rho$ modes. 
\end{abstract}

\maketitle

%\section{Introduction}
%%%%%%%%%%%%%%%%%%%%%%%%%%%%%%%%%%%%%%%%%%%%%%%%%
The main source of uncertainty in the extraction of the 
CKM matrix elements from the simple decay processes is our incomplete
knowledge of the non-perturbative  
dynamics that is necessary to compute the relevant 
hadronic matrix elements. In particular, to extract 
$\vert V_{cs} \vert$ and $\vert V_{ub} \vert$ from experimentally 
measured $D\to K^\ast \ell \nu$  and $B\to \rho \ell \nu$ decay rates
requires a
reliable QCD based computation of the following matrix elements: 
\bea
&&\hspace*{-6mm}\langle V (p^\prime,\varepsilon_\lambda) 
\vert \bar q \gamma_\mu Q \vert H(p)\rangle =
i \epsilon_{\mu \nu \alpha \beta} \varepsilon^{\ast \nu} 
p^\alpha p^{\prime\beta} {2 V(q^2)\over m_H +m_V},\cr
&&\hspace*{-6mm}\langle V (p^\prime,\varepsilon_\lambda) \vert  
\bar q \gamma_\mu \gamma_5 Q \vert H(p)
\rangle = 
{2 m_V (\varepsilon^\ast\cdot
q)\over q^2} A_0(q^2) q_\mu +\cr
&& (m_H + m_V) A_1(q^2) \left( \varepsilon_\mu^{\ast} - {\varepsilon^\ast\cdot
q\over q^2}q_\mu \right) -\cr
&&  A_2(q^2) { \varepsilon^\ast\cdot
q\over m_H + m_V} \left[ (p+p^\prime)_\mu - 
{m_H^2-m_V^2\over q^2} q_\mu \right], \nonumber
\eea
where we consider a generic $H\to V \ell \nu$ decay and use the standard 
decomposition in terms of four Lorentz invariant form factors, $V$,
$A_{1,2,0}$,  
which depend on $q^2 = (p-p^\prime)^2$. At $q^2=0$,
the axial form factors satisfy 
\bea\label{constr}
2 m_V A_0 (0) &=& (m_H + m_V) A_1(0)\cr
&&- (m_H - m_V) A_2(0)\;. 
\eea
We compute the above matrix elements on the lattice using the complete 
${\cal O}(a)$ non-perturbatively improved Wilson quark action and operators, 
working in the quenched approximation. We generated two sets of 
independent gauge field configurations: $200$ on $24^3 \times 64$ lattice 
at $\beta = 6.2$ ($a^{-1}= 2.7(1)$~GeV), and $100$ on $32^3 \times 70$ at 
$\beta = 6.45$ ($a^{-1}= 3.7(1)$~GeV). 

\begin{figure}
 \centerline{\psfig{file=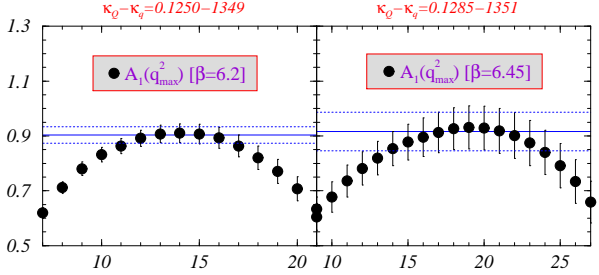,width=7.8cm,angle=0}}
\vspace*{-1.0cm}
\caption{\footnotesize Plateau of the form factor $A_1$, the function of the time $t$, 
between two source operators (fixed at $0$ and $t_F$, where $t_F=27$ at $\beta=6.2$ and 
$t_F=34$ at $\beta=6.45$). 
Illustration provided for $q^2_{max}=(m_H - m_V)^2$, with $m_H\simeq 1.8$~GeV and $m_V\simeq 1$~GeV.}
\label{fig1}
\end{figure}

We compute the following two- and three-point correlation functions:
\bea
&&\hspace*{-6mm}C^{(2)}_V(t) = \bigl<\sum_{\vec x} e^{i(\vec p -\vec q)\vec x} 
\left(\bar q \gamma_{\mu} q \right)^\dagger_{0} \left(\bar q \gamma_{\mu} q \right)_{\vec x, t } 
\bigr>,\nonumber \cr
&&\hspace*{-6mm}C^{(2)}_H( t) =\bigl< \sum_{\vec x} e^{i \vec p \vec x} 
\left(\bar Q \gamma_{5} q \right)^\dagger_{0} \left(\bar Q \gamma_{5} 
q \right)_{\vec x,  t } 
\biggr> ,\nonumber\cr
&&\hspace*{-6mm}C^{(3)}_{\mu \alpha}(t) = \nonumber\cr
&&\biggl<  
\sum_{\vec x, \vec y} e^{i\vec p \vec x - i \vec q \vec y}
\left(  \bar q \gamma_5 Q \right)^\dagger_{\vec x, t_{F}} 
{\left(\bar q \gamma_{\mu} Q \right)}_{\vec y, t} 
\left(\bar q \gamma_{\alpha} q \right)_{0}  \biggr>.  \nonumber
\eea
At $\beta=6.2$ we have 3 light ($q$) and $4$ heavy ($Q$) quark masses, 
whereas at $\beta=6.45$ we work with $4$ light and $6$ heavy quarks. 
The directly simulated vector mesons lie in the range $m_V\in (0.9, 
1.1)$~GeV, which means that the $K^\ast$-meson is within the grasp 
of our lattice study, while for the $\rho$-meson an extrapolation is needed. 
As for the heavy-light mesons, after sending the light quark mass (linearly) 
to zero, their masses at $\beta = 6.2$ are $m_{H_d}\in (1.7, 2.6)$~GeV. 
On the finer lattice ($\beta = 6.45$) that interval extends to $m_{H_d}\in (1.7, 
3.6)$~GeV. In other words, the $charm$ sector is simulated directly, while 
the $beauty$ can be reached through an extrapolation (a normal feature of 
current lattice studies in which fully relativistic heavy quarks are used). 
The form factors are extracted from the ratios 
\bea \label{ratio}
&& R_{\mu \alpha }(t) 
 = {C^{(3)}_{\mu \alpha} (t) \sqrt{{\cal Z}_V {\cal Z}_H}\over 
 C^{(2)}_V(t)C^{(2)}_H( t_F - t)}\cr &&
\hspace*{5mm}\stackrel{t_F\,\gg t\,\gg 0}{\Longrightarrow} {
\langle V(p^\prime) \vert \bar q \gamma_{\mu}  Q \vert
H(p) \rangle} \,. 
\eea 
To study the functional dependence of the form factors on $q^2$, we also consider 
$7$ different combinations of three-momenta for the interacting hadrons 
(for more details, please see ref.~\cite{prepa}). Discrete symmetries have been used 
to average over the equivalent momentum configurations. 
A suitable kinematical situation for a comparison of the lattice data at two 
values of the lattice spacing is when both mesons are at rest 
({\it i.e.} at $q^2=q^2_{max}$), because only the form factor $A_1$ contributes.
In fig.~\ref{fig1} we plot the signal for $A_1(q^2_{max})$ as extracted from the ratio~(\ref{ratio}) 
for both of our lattice spacings and for almost the same masses of mesons (in physical units). 
From this exercise we see that there are no large discretisation artefacts,
 when $\vert \vec p \vert= \vert \vec p^\prime \vert= 0$. 

For each fixed heavy quark mass and combination of momenta $\vec q$ and $\vec
p^\prime$, the leading dependence of each $F\in (V, A_1,A_2, A_0)$ on
the light final vector meson mass is expected
and seen to be linear, {\it i.e.}
\bea
&& F(m_V) = \alpha + \beta m_V\,,\nonumber
\eea
Fitting to this give the form factors for the transitions $H\to K^\ast$ and $H\to \rho$.  
The discussion of the dependence on the heavy quark (meson) mass is tightly related to
the $q^2$-shapes of the form factors. We chose to fit our (directly computed) form factors
to the pole/dipole forms
\bea \label{vmd}
 V(q^2)={V(0)\over (1 - q^2/M^2_{V})^2}  ,
&&A_1(q^2)={A_1(0) \over (1 - q^2/M^2_{1})},\nonumber\\
&& \\
  A_2(q^2)={A_{2}(0)\over (1 - q^2/M^2_{2})^2},  
&&A_{0}(q^2)={A_{0}(0) \over (1 - q^2/M^2_{0})^2},\nonumber
\eea
additionally constrained by the condition~(\ref{constr}). The pole/dipole forms~(\ref{vmd})
reconcile the $t$-channel pole dominance with the HQET scaling laws,  
according to which (for small recoil momenta) the form factor $A_1$ ($V$, $A_{2,0}$) multiplied 
by ${m_H}^{+1/2}$ (${m_H}^{-1/2}$) scales as a constant, up to $1/m_H$ corrections~\cite{isgur-wise}. 
In addition, the forms~(\ref{vmd}) are consistent with the ``$m^{-3/2}$" 
scaling law arising in the limit in which the light meson is very energetic 
(LEL)~\cite{LEET}.

\begin{figure}
 \centerline{\psfig{file=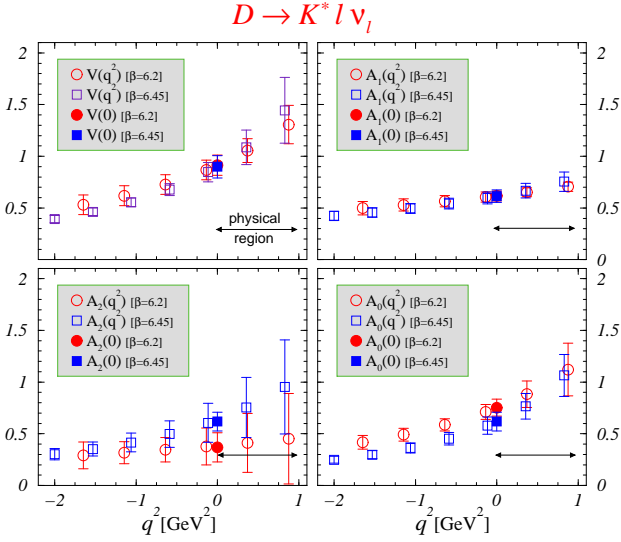,width=1.1\hsize,angle=0}}
\vspace*{-1.0cm}
\caption{\footnotesize Form factors for the $D\to K^\ast$ transition from both 
sets of our lattice data. Physical region for $D\to K^\ast \ell \nu$ is 
$0 \leq q^2\leq  0.95 \ \gev^2$. Note that $A_2(0)$ is obtained from $A_1(0)$ 
and $A_0(0)$ by using
eq.~(\ref{constr}).}
\label{fig2}
\end{figure}

With the $H\to K^\ast$ and $H\to \rho$ form factors fitted to the 
pole/dipole forms~(\ref{vmd}) we can interpolate  in 
the inverse heavy meson mass to reach $m_H=m_D$, 
and extrapolate to $m_H=m_B$. To that end, we use the
HQET scaling laws, and for a fixed value of  $v\cdot p^\prime = (m_H^2 + 
m_{K^\ast}^2 - q^2)/(2 m_H)$, we fit our data to 
\bea \label{extrap}
&&F(v\cdot p^\prime) m_H^{d/2} = a + b/m_H + c/m_H^2\;. 
\eea
where $d=+1$ for $F=A_1$, and $d=-1$ otherwise. The difference between 
this form and the linear one ($ c = 0$) is used to estimate the systematic 
error (as in ref.~\cite{pion}). Such a difference is completely
negligible in the case of $D$-meson because $m_D$ is very close 
to the lightest of the heavy-light mesons directly simulated on the lattice. 
$D\to K^\ast$ transition form factors are shown in fig.~\ref{fig2},
and the result of the fit to the forms~(\ref{vmd}) is given in 
tab.~\ref{tab1}. 
\begin{table}[b]
\caption{\footnotesize Result of the fit  of the $D\to K^\ast \ell \nu$ 
form factors to eq.~(\ref{vmd}). }
\label{tab1}
\hspace*{-4mm}\begin{tabular}{|c|cc|cc|}\cline{2-5}
\multicolumn{1}{c|}{}&\multicolumn{2}{c|}
{$\beta=6.2$}& 
\multicolumn{2}{c|}{$\beta=6.45$}\\
\hline
$F$   & $F (0)$ & $M_F [\gev]$ 
      & $F (0)$ & $M_F [\gev] $ 
\\
\hline
$V$    & 0.91(10) & 2.3(3)  & 0.90(11) & 2.0(2)\\
$A_1$  & 0.62(5)  & 2.6(5)  & 0.61(6)  & 2.2(3)\\
$A_2$  & 0.37(14) & 3(23)   & 0.64(14) & 2(1) \\ \hline
\end{tabular}
\end{table}
We also remind the reader that $A_0$ does not enter the expression for the 
decay rate (see eg.~\cite{LCSR}). $A_2$, instead, enters the part describing 
the longitudinally polarised vector meson. Since the quality of our signals for $A_2$ 
is low (much worse than for $A_0$), we use the exact relation~(\ref{constr}) 
to compute $A_2(0)$. 
From the results of tab.~\ref{tab1}, for the integrated decay rate, we 
obtain
\bea
\vert V_{cs}\vert^{-2} \Gamma(D^- \to K^{\ast 0} \ell \nu)=
0.066(14)\ ps^{-1}_{[\beta
= 6.2]} ,\nonumber \\
\hspace*{6mm} 0.062(15)\ ps^{-1}_{ [\beta = 6.45]},\nonumber
\eea
which after comparison to the recently measured branching ratio~\cite{cleo} lead
to $\vert V_{cs} \vert = 0.99(9)$ and $\vert V_{cs}\vert  = 1.03(12)$, respectively. 

An additional comparison with the experimental data is 
provided for the ratios of the form factors at $q^2=0$. Our results
\bea
&&V/A_1 = 1.48(12)_{\beta=6.2},\,1.46(11)_{\beta=6.45},\nonumber \\
&&A_2/A_1 = 0.6(3)_{\beta=6.2},\,1.0(2)_{\beta=6.45},\nonumber 
\eea
agree very well with $(V/A_1)^{exp.} = 1.50(7)$~\cite{focus}. The agreement
with  $(A_2/A_1)^{exp.} = 0.88(9)$~\cite{focus} is only marginal, as
discussed above.
\begin{figure}[b!]
 \centerline{\psfig{file=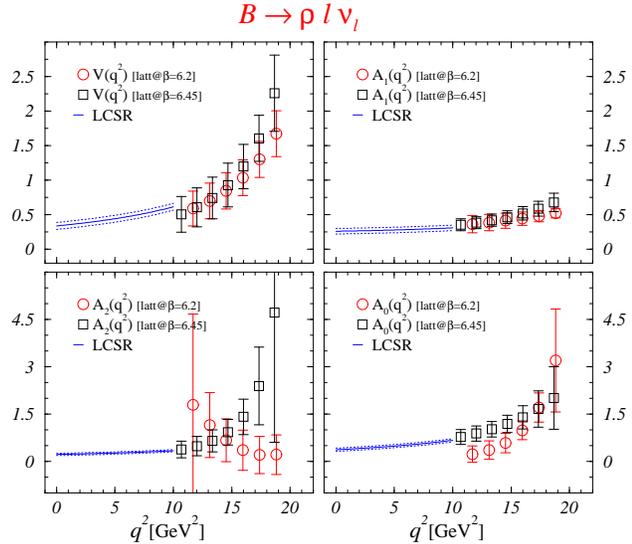,width=1.1\hsize,angle=0}}
\vspace*{-1.0cm}
\caption{\footnotesize $B\to \rho$ form factors on our lattices are accessed for 
$q^2 > 10\ \gev^2$. For the decay rate they are combined with the
lightcone sumrule (LCSR) results 
in the $q^2 < 10\ \gev^2$ region (which are also shown in the figure).}
\label{fig3}
\end{figure}

We next discuss the $B\to \rho$ form factors. The results of (quadratic) 
extrapolation~(\ref{extrap}), for each fixed $(v\cdot p^\prime)$, are shown 
in fig.~\ref{fig3}. We observe the standard effect that after the heavy quark 
extrapolation the form factors fall into the region 
$q^2 > 10\ \gev^2$. 
As compared to the benchmark calculation by UKQCD~\cite{ukqcd}, our results 
have larger errors (in spite of the fact that our statistics is higher). 
In particular, the errors for our $A_2^{B\to \rho}(q^2)$ are of ${\cal
O}(100\% )$. 

To compute the decay rate we have to integrate over the entire phase
space. Therefore we combine the lattice results for
$(d\Gamma/dq^2)_{q^2 > 10 \gev^2}$ with lightcone sumrule
results for $(d\Gamma/dq^2)_{q^2 < 10 \gev^2}$~\cite{LCSR} (which are
expected to be reliable for low values of $q^2$).

To obtain $(d\Gamma/dq^2)_{q^2 > 10 \gev^2}$, we use the pole/dipole
forms~(\ref{vmd}), eliminating $F(0)$ in favour of one of our points
$F(q_0^2)$. We choose $q_0^2 = 14.6 \gev^2$ in the middle of the region
covered by our results and fit to
\bea
&&{ F(q^2) \over F(q_0^2)} = \left( { 1 - q_0^2/M_F^2 \over 1 - q^2/M_F^2  }\right)^p\,,
\eea
where $p=1$ for the case $F=A_1$, and $p=2$ otherwise. 
\begin{table}[tb]
\caption{\footnotesize Result of the fit of the $B\to \rho \ell \nu$ form factors (see the text) 
$q^2_0=14.53\ \gev^2$ for the data at $\beta=6.2$, and $q^2_0=14.68\ \gev^2$ at  $\beta=6.45$.}
\label{tab2}
\hspace*{-4mm}\begin{tabular}{|c|cc|cc|}\cline{2-5}
\multicolumn{1}{c|}{}&\multicolumn{2}{c|}
{$\beta=6.2$}& 
\multicolumn{2}{c|}{$\beta=6.45$}\\
\hline
$F$   & $F (q_0^2)$ & $M_F [\gev]$ 
      & $F (q_0^2)$ & $M_F [\gev] $ \\
\hline
$V$    & 0.84(26)  & 5.4(5)  & 0.93(31) & 5.2(4)\\
$A_1$  & 0.41(11)  & 5.9(1.2)  & 0.46(9)  & 5.3(4)\\
$A_2$  & 0.7(7)    & --     & 0.9(7)   & -- \\ \hline
\end{tabular}
\end{table}
Results of this single parameter interpolation procedure are listed in
tab.~\ref{tab2}. Notice that we neglect the slope of $A_2$, for which
a flat $q^2$-form with ${\cal O}(100\%)$ of error on the central value
should be conservative enough. We finally obtain
\bea
 \vert V_{ub}\vert^{-2} \Gamma(\bar B^0\to \rho^+ \ell \nu) = (17\pm 3) \ ps^{-1}_{[\beta
= 6.2]} \,,\nonumber \\
 \hspace*{6mm} (19\pm 4) \ ps^{-1}_{[\beta
= 6.45]}\,,\nonumber 
\eea
which we then match with the experimental branching ratio (as measured by CLEO, 
BaBar and Belle~\cite{exp-brho}) to extract $\vert V_{ub}\vert$. We find 
\bea
&&\vert V_{ub}\vert = 0.0034(6)\, ,\nonumber 
\eea
where we added all the errors in quadrature.

\begin{figure}[h!]
 \centerline{\psfig{file=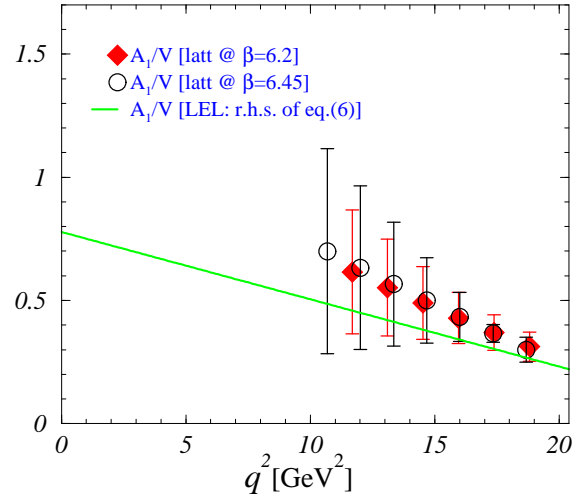,width=7.5cm,angle=0}}
\vspace*{-1.0cm}
\caption{\footnotesize Ratio of $B\to \rho$ form factors, $A_1(q^2)/V(q^2)$, computed on the lattice 
(at large $q^2$'s is plotted against the r.h.s. of eq.~(\ref{lel}) denoted by a full line).}
\label{fig4}
\end{figure}

As a final exercise, we check the relation among form factors which  
holds true in the LEL (for a recent discussion see refs.~\cite{LEET,SCET}), namely 
\bea\label{lel}
&&{A_1(q^2)\over V(q^2)} = {2 E_\rho m_B \over  (m_B + m_\rho)^2} \;. 
\eea
This relation is verified in the LCSR approach (note that $E_\rho \simeq m_B/2$ 
for $q^2\approx 0$ in the  $B$-meson rest frame). In fig.~\ref{fig4}, we plot  
the ratio of our $B\to \rho$ form factors (computed on the lattice), and compare 
them to the r.h.s. of eq.~(\ref{lel}). Interestingly, 
we do not see deviations from that relation~(\ref{lel}) 
in spite of the fact that the lattice results are produced at 
large $q^2$. 


\begin{thebibliography}{99}

\bibitem{prepa}
A.~Abada  {\it et al.} [SPQcdR], in preparation.

\bibitem{isgur-wise}
N.~Isgur and M.~B.~Wise, Phys.\ Rev.\ D {\bf 42} (1990) 2388.
%%CITATION = PHRVA,D42,2388;%%

\bibitem{LEET}
J.~Charles {\it et al.}, Phys.\ Rev.\ D {\bf 60} (1999) 014001,{\tt 
[hep-ph/9812358]}.
%%CITATION = HEP-PH 9812358;%%


\bibitem{pion}
A.~Abada {\it et al.} [APE], 
Nucl.\ Phys.\ B {\bf 619} (2001) 565,
{\tt [hep-lat/0011065]};
%%CITATION = HEP-LAT 0011065;%%
K.~C.~Bowler {\it et al.}  [UKQCD],
Phys.\ Lett.\ B {\bf 486} (2000) 111, {\tt 
[hep-lat/9911011]}.
%%CITATION = HEP-LAT 9911011;%%

\bibitem{LCSR}
P.~Ball and V.~M.~Braun,
Phys.\ Rev.\ D {\bf 58} (1998) 094016,{\tt [hep-ph/9805422]}.
%%CITATION = HEP-PH 9805422;%%

\bibitem{focus}
J.~M.~Link {\it et al.}  [FOCUS],
{\tt [hep-ex/0207049]}.
%%CITATION = HEP-EX 0207049;%%


\bibitem{cleo}
G.~Brandenburg {\it et al.}  [CLEO],
{\tt [hep-ex/0203030]}.
%%CITATION = HEP-EX 0203030;%%


\bibitem{ukqcd}
J.~M.~Flynn {\it et al.}  [UKQCD],
Nucl.\ Phys.\ B {\bf 461} (1996) 327,
{\tt [hep-ph/9506398]}.
%%CITATION = HEP-PH 9506398;%%

\bibitem{exp-brho}
B.~H.~Behrens {\it et al.}  [CLEO],
Phys.\ Rev.\ D {\bf 61} (2000) 052001,
{\tt [hep-ex/9905056]}; 
%%CITATION = HEP-EX 9905056;%%
B.~Aubert {\it et al.}  [BABAR],
{\tt [hep-ex/0207080]},
%%CITATION = HEP-EX 0207080;%%
Y. Kwon  [Belle], talk presented at ICHEP 2002, Amsterdam.


\bibitem{SCET}
C.~W.~Bauer {\it et al.}, Phys.\ Rev.\ D {\bf 63} (2001) 114020
{\tt [hep-ph/0011336]},
%%CITATION = HEP-PH 0011336;%%
G.~Burdman and G.~Hiller,
Phys.\ Rev.\ D {\bf 63} (2001) 113008,
{\tt [hep-ph/0011266]},
%%CITATION = HEP-PH 0011266;%%
M.~Beneke{\it et al.}, {\tt [hep-ph/0206152]}.
%%CITATION = HEP-PH 0206152;%%

\end{thebibliography}
\end{document}